\documentstyle[11pt,newpasp,twoside,epsf]{article}
\def\edcomment#1{\iffalse\marginpar{\raggedright\sl#1\/}\else\relax\fi}
\reversemarginpar

\begin{document}
\title{Near-IR Imaging of LSB Galaxies: the High and Low HI Content Cases}
 \author{Gaspar Galaz\altaffilmark{1}}
\affil{Carnegie Observatories. Las Campanas Observatory. Casilla 601, La Serena,
Chile. E-mail: gaspar@azul.lco.cl}
\altaffiltext{1}{Andes-Carnegie Fellow}

\begin{abstract}
Recently acquired near-IR imaging ($J$ and $K_s$) of low 
surface brightness galaxies is presented. 
The data includes 52 galaxies with log[M$_{HI}$/M$_{\odot}$] $\le$ 9.0 
and 58 galaxies with log[M$_{HI}$/M$_{\odot}$] $\ge$ 9.5. These
galaxies have been selected from the catalogue of Impey et al. (1996), and all of them 
are observable from both hemispheres. The principal goal of this research is
to investigate the poorly understood giant and old stellar content of these galaxies.
Current work includes total and isophotal photometry, and comparison with 
spectrophotometric models of galaxy evolution, including the role of 
age and metallicity. Already allocated observing time at LCO for this year will 
provide high S/N $B$ and $I$ imaging.
\end{abstract}

\section{Introduction: Background and Rationale}

Low surface brightness galaxies (LSBGs) have been the subject of interest 
since Zwicky (1957) and Disney (1976)
emphasized the fact that the central surface brightnesses of disk galaxies in the 
Hubble sequence Sa-Sb-Sc (which fall in a rather narrow range) might be the result 
of a selection effect, which is due to the difficulty in discovering galaxies 
of very low surface brightness. In practice, LSB means galaxies whose 
central surface brightness is fainter than 22.0 mag arcsec$^{-2}$ in the
$B$ band, and seem to have different properties than those of brighter
galaxies, which might imply a different evolutionary track.

Previous studies (Longmore et al. 1982) has shown that LSBGs are
much more gas rich than high surface brightness ones, and bluer than ``normal''
late type galaxies (de Blok, van der Hulst \& Bothun 1995; 
Sprayberry et al. 1995; Dalcanton, Spergel \& Summers 1997).
These facts combined with their low metallicity content (McGaugh 1994; 
de Blok \& McGaugh 1997), result in a low star formation rate in an unevolved system.
However, because of the low surface brightness of the 
underlying population, only a small fraction of the total number of stars is
needed to make the colors significantly blue. Therefore, the study of the M/L 
ratio in the near-IR for these galaxies would lead to a clearer picture. 
Also, more
observational constraints are needed in order to explain the blue colors 
and low metallicity and to disentangle the role played by age and 
metallicity. Combined near-IR and optical observations are a 
perfect probe to clarify these issues.

Knezek \& Wroten (1994) observed a small sample of LSBGs in $J$, $H$ and $K$, but
the sample was very small and biased towards massive galaxies. 

The fact that the integrated colors of some LSBGs are quite red 
(like Malin-type objects) suggests that they contain a 
significant old population of stars and 
were therefore not formed recently. The 
near-IR colors are good tracers {\em of the mass}, where additionally 
the evolution factor is minimized by the 
weakly evolving dominant population of red giant stars and low mass stars.
Moreover, integrated color observations by McGaugh (1992) show that only
the reddest LSBGs have bulges. The super-giant galaxies Malin 1 and 
Malin 2 appear to be redder than the majority of LSBGs with luminosities
that are comparable to those of normal spirals. Our preliminary 
results in the near-IR seem to agree with this picture, in the sense that more massive
LSBGs (high HI mass) are redder than less massive (small HI masses), low luminosity
LSBGs. 

With these caveats in mind, this systematic study will provide
fundamental properties of LSBGs both in the optical and in the
near-IR (for example, to derive the size of these galaxies in terms of the 
surface brightness in both the near-IR and in the optical). 
This offers the possibility of tracing accurately the mass
distribution using near-IR imaging, and the 
luminosity distribution using optical colors, helping to trace
better the evolutionary tracks of these galaxies. 
The spread in the observed integrated 
optical colors of LSBGs is large, showing that such objects have 
had a wide range of evolutionary histories. Near-IR colors prevent the melting
of the light of all the star populations to infer the evolutionary 
(and metallicity) path of these galaxies. 
Optical and near-IR 
colors combined give a complete picture of the spatial distribution of the 
young and old stellar populations, as well as allow to quantify the 
degree of internal absorption 
by dust. Moreover, recent modeling of LSBGs predict that, from 
the assumption that blue LSBGs are currently 
undergoing a period of enhanced star formation, there should exist
a population of red, non-bursting, quiescent LSBGs
(Gerritsen \& de Blok 1999; Bell et al. 2000; Beijersbergen, 
de Blok, \& van der Hulst 1999). 
These galaxies should then also be metal-poor and 
gas-rich, and share many of the properties 
of the LSBGs observed in the optical. Near-IR imaging is a 
valuable window to test this last hypothesis and to study the 
stellar content of LSBGs, allowing to compare the
properties of their old stellar population with the features observed for the 
young populations using optical data, particularly from blue bands. 

\section{The Sample}

In order to investigate these issues, I have selected two 
subsamples of LSBGs from the catalogue of Impey et al. (1996). The two subsamples differ 
in the HI mass of their galaxies. The first subsample contains LSBGs with 
log[M$_{HI}$/M$_{\odot}$] $\le$ 9.0 (52 galaxies, {\bf sample A})
and the second subsample galaxies with log[M$_{HI}$/M$_{\odot}$] $\ge$ 9.5
(58 galaxies, {\bf sample B}). In the final catalogue, {\em only} 
galaxies with $\mu_0(B) \le 23.5$ mag arcsec$^{-2}$ are included. 
This $B$ limiting central magnitude 
corresponds roughly to the equivalent limiting magnitude in $K$ for the 
IR Classic Camera at the Las Campanas 2.5m du Pont telescope for a stellar image in one hour 
integration time (5$\sigma$ and 1.5 arcsec aperture), 
given that $<B-K> \sim 2.5-3.0$ for LSBGs (see for example Tully \& 
Verheijen 1997). All the galaxies already have relevant 
information, such as radial velocities (i.e. distance),  HI content, 
$B_{total}$ magnitudes, and the central blue surface magnitude $\mu_{0}$. Also 
some morphological information is available (Hubble types). 

\section{Status of the Project and Prospects}

Currently, most of the near-IR observations are completed and 90\%
of them are reduced. All the sample will be imaged in $B$ and $I$ 
during 2000. Therefore, the final catalogue will include high S/N
($\sim 10-15$) $B$, $I$, $J$ and $K_s$ imaging of 110 LSBGs from the catalogue of
Impey et al. (1996), as described in \S 2. The current work with the 
data includes (1) structural characterization of the near-IR images (sizes,
shapes and morphology), (2) computation of different kind of calibrated magnitudes
for the whole sample, (3) characterization of the photometric profiles 
in terms of magnitudes and colors, and (4), comparison of derived photometric results with 
spectrophotometric models of galaxy evolution, considering the role of metallicity and dust,
e.g. using PEGASE (Fioc \& Rocca-Volmerange 1997) and GISSEL96 (Charlot, Worthey, \& 
Bressan 1996). 
The approach will be similar to that applied by Galaz (2000) for a sample of
E+A galaxies and by Bell et al. (2000). 
I emphasize that this is currently a unique large
and homogeneous database of LSBGs with reduced near-IR imaging.  

Figures 1 and 2 show $J$ and $K_s$ mosaiced images of 38 reduced
LSBGs from sample A. Figures 3 and 4 mosaiced images of 47 reduced
LSBGs from sample B. 
\begin{figure}
\plottwo{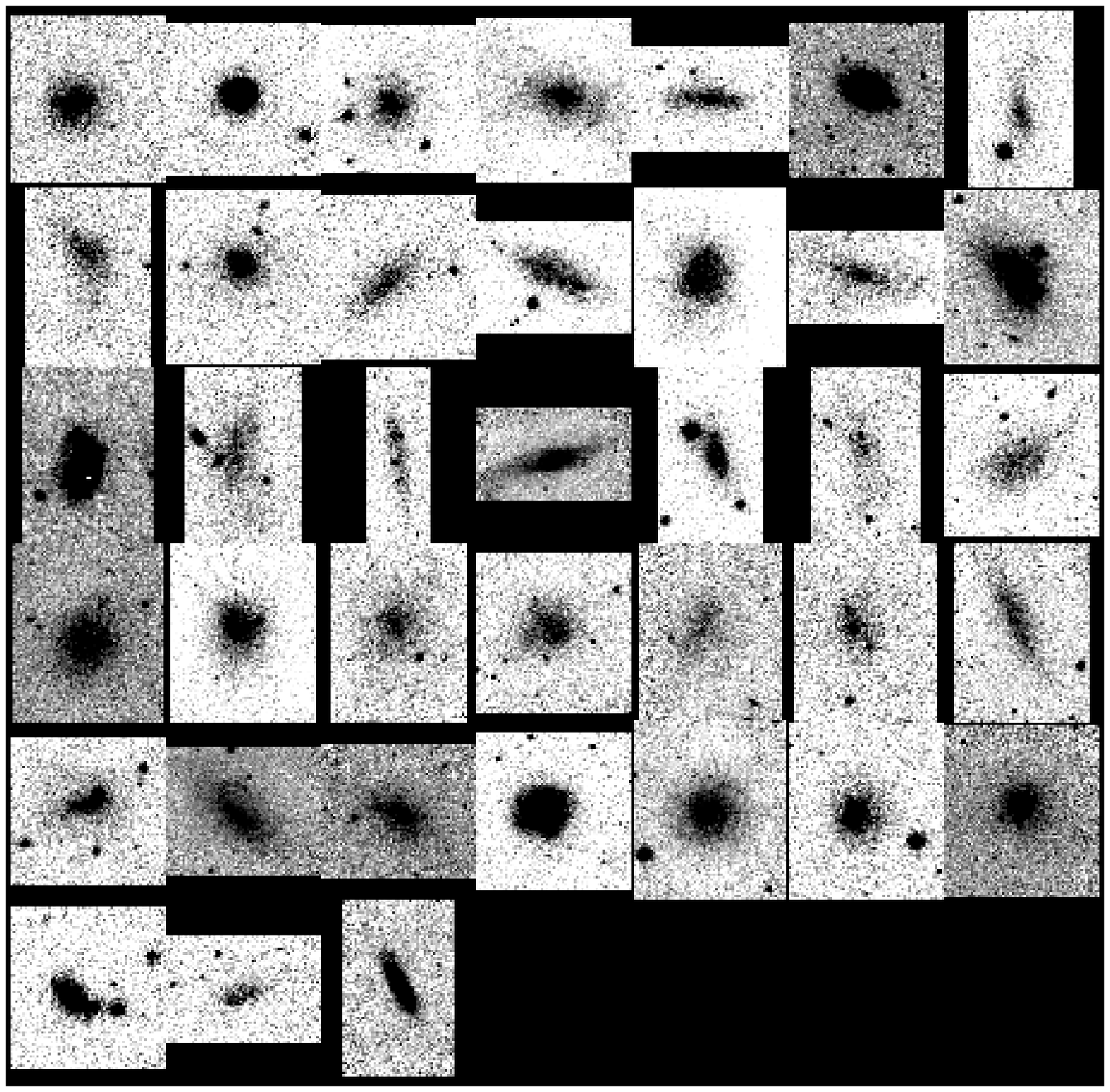}{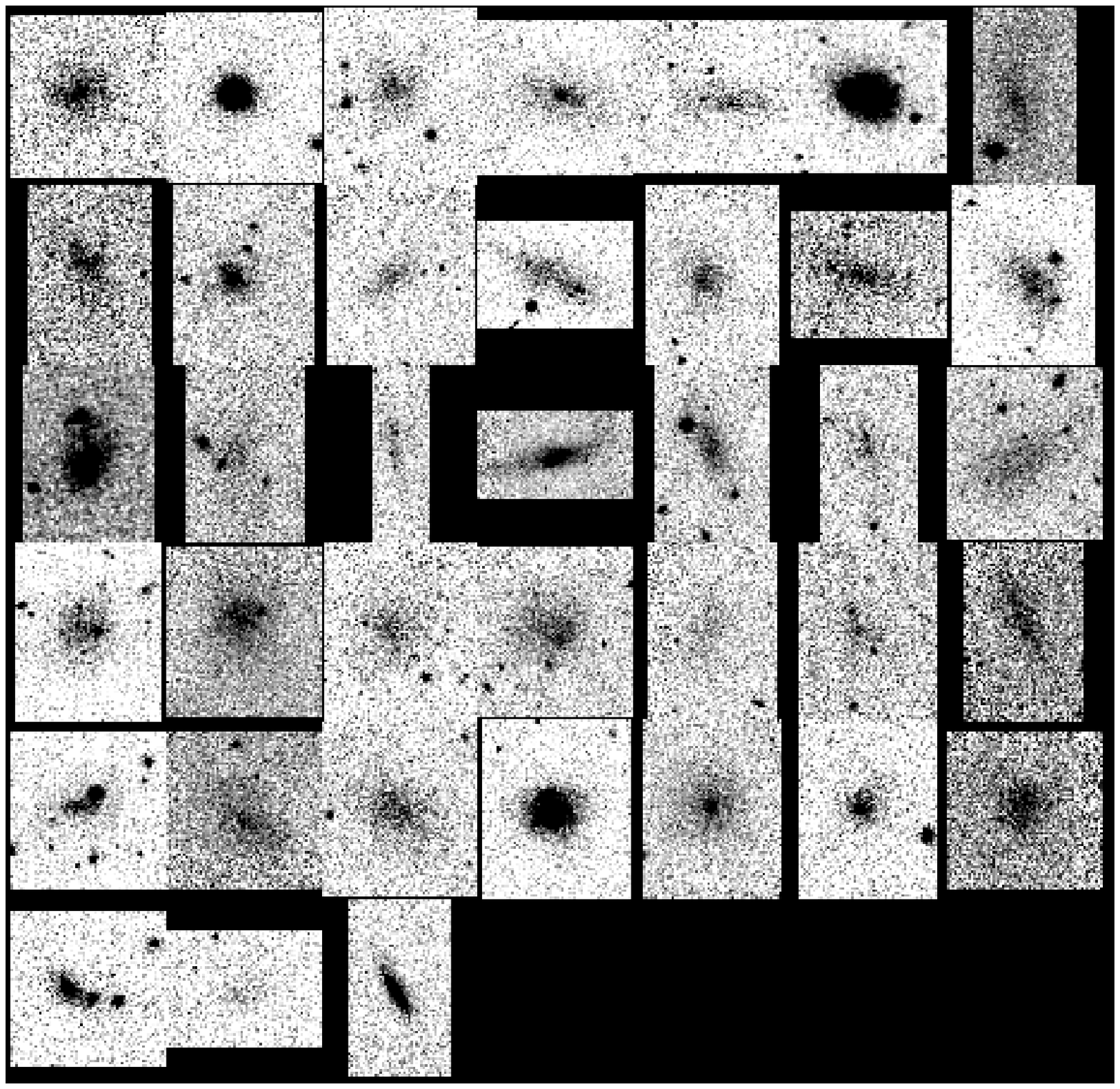}
\caption{Mosaic of 38 low surface brightness galaxies of {\bf sample A} already reduced
in the near-IR. In left panel $J$ images and in right panel $K_s$ images. 
Typical frame size is $\sim 1.5^\prime \times 1.5^\prime$. Total exposure time for
each galaxy varies between 30 min and 70 min in $J$, and between 
45 min and 110 min in $K_s$.}
\end{figure}
\begin{figure}
\plottwo{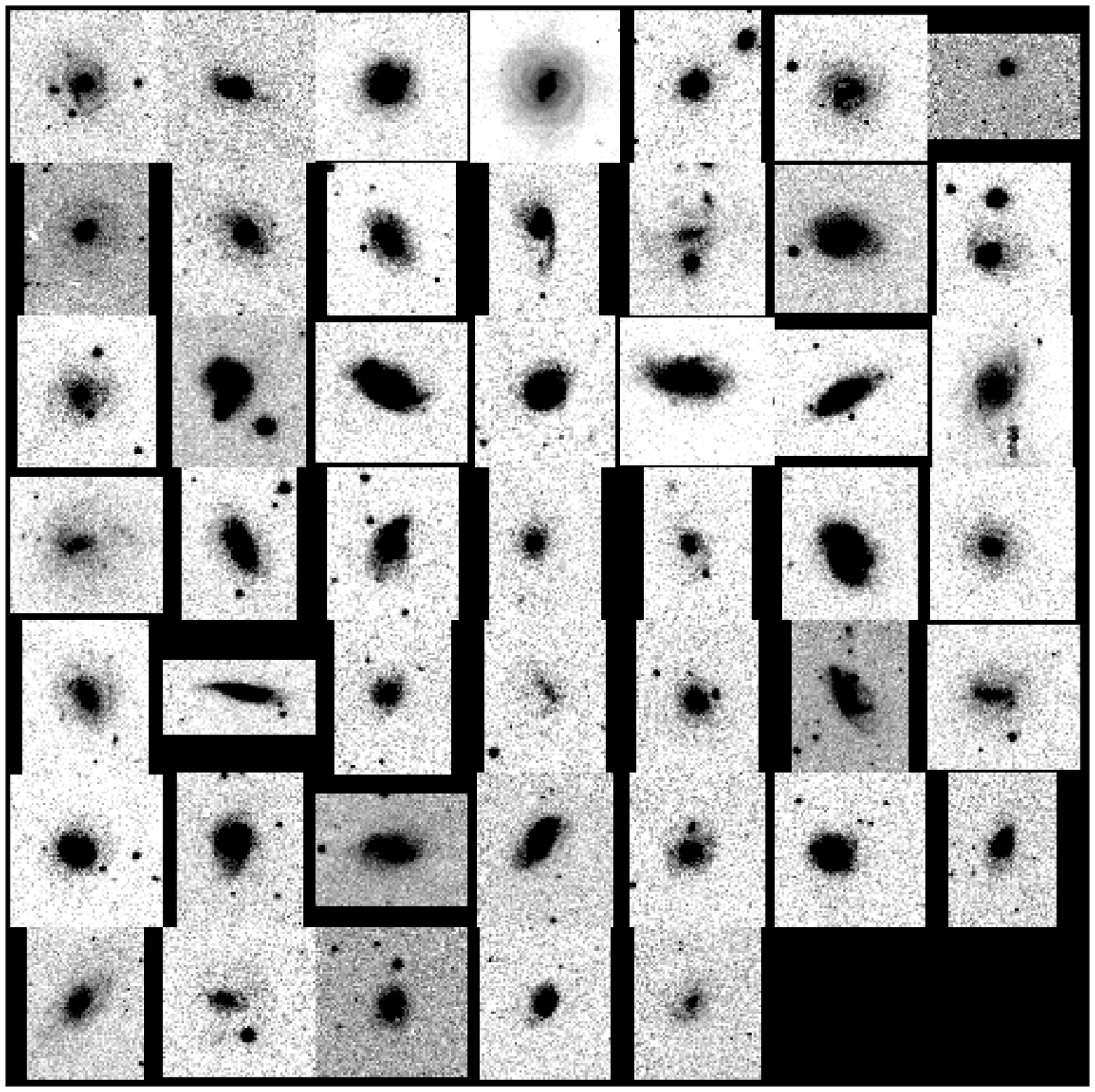}{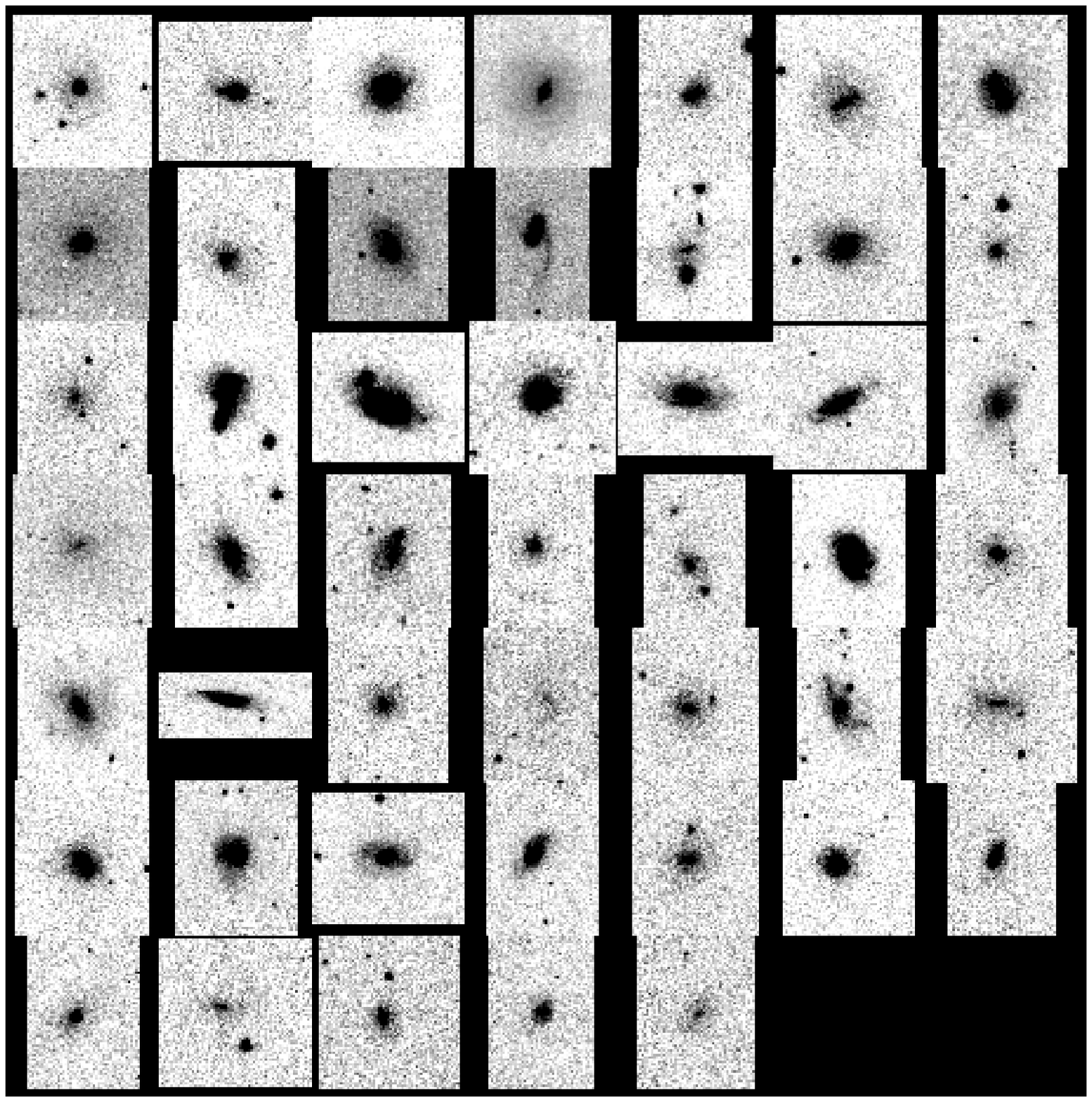}
\caption{Similar to Fig. 1 but for {\bf sample B}.}
\end{figure}
The core of the reduction pipeline uses
a modified version of DIMSUM,
and has been applied successfully to all the data (see also Galaz 2000). 
In the future, we expect to observe fainter LSBGs in the near-IR and in the 
optical, using larger telescopes. 

\acknowledgements

I acknowledge Carnegie Observatories for the huge (but necessary) 
telescope time allocated for this project. This work is supported under
agreement between Fundaci\'on Andes and Carnegie Institution of 
Washington.

\end{document}